%Paper: hep-th/9405021
%From: R. A. Sharipov <root@bgua.bashkiria.su>
%Date: Mon, 3 May 93 15:03:59 +0600

% Typeset by AmS-TeX, version 2.1,
\input amstex
\documentstyle{amsppt} %this line is optional
\pagewidth{18cm}\pageheight{23cm}
\loadbold
\CenteredTagsOnSplits
\rightheadtext{Dynamical Systems accepting the normal shift\dots}
\topmatter
\title
Dynamical systems accepting the normal shift\\
on an arbitrary Riemannian manifold.
\endtitle
\author
Boldin A.Yu., Dmitrieva V.V., Safin S.S., Sharipov R.A.
\endauthor
\address
Department of Mathematics, Bashkir State University,
Frunze street 32, 450074 Ufa, Russia.
\endaddress
\email
root\@bgua.bashkiria.su
\endemail
\date
August 6, 1993
\enddate
\abstract
     Newtonian dynamical systems which accept the normal shift
on an arbitrary Riemannian manifold are considered. For them the
determinating equations making the weak normality condition are
derived. The expansion for the algebra of tensor fields is
constructed.
\endabstract
\endtopmatter
\document
\head
1. Introduction.
\endhead
     The concept of dynamical systems  accepting the normal shift
was introduced in \cite{1} (see also \cite{2}\footnotemark). It
appears as the result of transferring the classical Bonnet
transformation from geometry to the field of dynamical systems.
In \cite{1} and \cite{2} the Euclidean situation was studied i.e.
the dynamical systems in $\Bbb R^n$ accepting the normal shift
(we refer the reader to that papers for the detailed
bibliography). \par
\footnotetext{See also {\bf chao-dyn/9403003}
and {\bf patt-sol/9404001}}%
     Apart from recent results of \cite{1} and \cite{2} one can
find another purely geometric generalization of Bonnet
transformation from \cite{3} and \cite{4} which is realized as a
normal shift along the geodesics on some Riemannian manifold. The
natural question here is how do these two generalizations relate
each other. This question was investigated in \cite{5}. There the
special subclass of dynamical systems accepting the normal shift
was separated for which the normal shift is equivalent to the
geometrical generalization of the Bonnet transformation for some
conformally-Euclidean metric in $\Bbb R^n$. Such systems are
called metrizable. Comparing the explicit description of
metrizable dynamical systems given in \cite{5} with the examples
of \cite{1} and \cite{2} one can conclude that non-metrizable
systems do exist. Therefore the the concept of normal shift
along the dynamical system is wider than the Bonnet transformation
for conformally-Euclidean metrics. However it do not embrace the
case of Bonnet transformation for arbitrary metric. In this paper
below we consider the most general situation and study the
dynamical systems accepting the normal shift on an arbitrary
Riemannian manifold.
\head
2. Newtonian dynamical systems on the Riemannian manifold.
\endhead
     The main object considered in \cite{1} and \cite{2} is the
second order dynamical system in $\Bbb R^n$
$$
\ddot\bold r=\bold F(\bold r,\dot\bold r)\tag2.1
$$
reflecting the Newton's second law. Now let $\bold r$ be not
radius-vector of a point in $\Bbb R^n$ but the vector of local
coordinates for some manifold $M$. This case let us write the
equations \thetag{2.1} in form of the system of the first order
equations
$$
\dot x^i=v^i \hskip 10em
\dot v^i=\Phi^i(\bold r,\bold v)\tag2.2
$$
Systems of differential equations of the form \thetag{2.2} are
traditionally connected with vector fields on manifolds. In this
particular case the right hand sides of \thetag{2.2} form the
components of the vector field not on $M$ however but on the
tangent bundle $TM$
$$
\boldsymbol\Phi=v^1 \frac{\partial}{\partial x^1}+\dots+
v^n \frac{\partial}{\partial x^n}+\varPhi^1
\frac{\partial}{\partial v^1}+\dots+
\varPhi^n \frac{\partial}{\partial v^n}\tag2.3
$$
First $n$ components of the vector field \thetag{2.3} separately
can be interpreted as components of the velocity vector
$$
\bold v = v^1 \frac{\partial}{\partial x^1}+\dots+
v^n \frac{\partial}{\partial x^n}\tag2.4
$$
tangent to $M$. Rest part of components in \thetag{2.3} do not
admit such interpretation. But if the manifold $M$ is equipped
with Riemannian metric $g_{ij}$ and with the metrical connection
$\varGamma^k_{ij}$ then  using the components of \thetag{2.3} one
can form the following quantities
$$
F^i=\varPhi^i+\varGamma^i_{jk} v^k v^j\tag2.5
$$
that do behave like the components of some vector tangent to $M$
when we change the local map on $M$
$$
\bold F =F^1 \frac{\partial}{\partial x^1}+\dots+
         F^n \frac{\partial}{\partial x^n}\tag2.6
$$
Same indices on the different levels in \thetag{2.5} and
everywhere below imply summation. Vector $\bold F$ from
\thetag{2.6} with the components of the form \thetag{2.5} is
natural to be considered as a vector of force. The analogy of
\thetag{2.1} and \thetag{2.2} then becomes complete. The existence
of the vector $\bold F$ lets us speak about the angle between force
and velocity. It lets us also break the force $\bold F$ into two
parts directed along the velocity and perpendicular to the
velocity. When $\bold F=0$ equations \thetag{2.2} become the
equations of geodesic line. \par
\head
3. Metric on the tangent bundle and the expansion
of the algebra of tensor fields.
\endhead
     Let us consider again the manifold $M$ with the Riemannian
metric $g_{ij}$. Vector fields on the tangent bundle $TM$ have
the form
$$
\bold V=X^1 \frac{\partial}{\partial x^1}+\dots+
X^n \frac{\partial}{\partial x^n}+
W^1 \frac{\partial}{\partial v^1}+\dots+
W^n \frac{\partial}{\partial v^n}\tag3.1
$$
First $n$ components of $\bold V$ in \thetag{3.1} like in
\thetag{2.4} are interpreted as the components of the vector
$\bold X$ tangent to $M$. The whole set of components of
\thetag{3.1} is transformed as follows
$$
\tilde X^i=\frac{\partial\tilde x^i}{\partial x^j}X^j \hskip 8em
\tilde W^i=\frac{\partial^2\tilde x^i}{\partial x^j \partial x^k}
v^k X^j+\frac{\partial\tilde x^i}{\partial x^s} W^s\tag3.2
$$
when one change the local variables $x^1,\dots, x^n$ for
$\tilde x^1,\dots,\tilde x^n$. The components of metric
connection $\varGamma^k_{ij}$ under the same action are
transformed as follows
$$
\tilde\varGamma^i_{pq} \frac{\partial\tilde x^p}{\partial x^j}
\frac{\partial\tilde x^q}{\partial x^k}=\frac{\partial\tilde x^i}
{\partial x^s} \varGamma^s_{jk} - \frac{\partial^2\tilde x^i}
{\partial x^j \partial x^k}\tag3.3
$$
Comparing \thetag{3.2} and \thetag{3.3} we conclude that the
following quantities $Z^i$ produced from the components of
\thetag{3.1}
$$
Z^i = W^i+\varGamma^i_{jk} v^k X^j \tag3.4
$$
are transformed like the components of tangent vector to $M$
under the change of local map on $M$. For the vector \thetag{2.3}
the corresponding vector \thetag{3.4} coincides with
\thetag{2.5}. So the vector $\bold V$ tangent to $TM$ can be
replaced by two vectors $\bold X$ and $\bold Z$ tangent to $M$.
This gives rise to the pair of linear maps $\pi$ and $\rho$ where
$\pi$ is canonical projection from the bundle $TM$ to the base
manifold $M$
$$
\bold X = \pi(\bold V)\hskip 10em \bold Z=\rho(\bold V)\tag3.5
$$
The relationship \thetag{3.4} lets us introduce the Riemannian
metric on the tangent bundle $TM$ by forming the following
quadratic form on the set of vectors \thetag{3.1}
$$
\tilde g(\bold V,\bold V) = g(\pi(V),\pi(V))+g(\rho(V),\rho(V))
\tag3.6
$$
In terms of differentials of local coordinates the metric
\thetag{3.6} is written as follows
$$
\tilde g = (g_{ij}+
\varGamma^p_{ir}v^rg_{pq}\varGamma^q_{js}v^s) dx^i
dx^j+2(g_{iq}\varGamma^q_{js}v^s)dx^idv^j+g_{ij}dv^idv^j
\tag3.7
$$
Metric tensor for \thetag{3.7} is determined by the initial
metric tensor $g_{ij}$ of $M$. Its matrix has the natural block
structure
$$
\spreadmatrixlines{0.5em}
\tilde g_{ij}=\left(\matrix\format \l &\quad\l\\
g_{ij}+\varGamma^p_{ir}v^rg_{pq}\varGamma^q_{js}v^s  &
\varGamma^p_{ir}v^rg_{pj}\\g_{iq}\varGamma^q_{js}v^s &
g_{ij}\endmatrix\right)\tag3.8
$$
Upper left block corresponds to the coordinates on the base while
the lower right block of \thetag{3.8} corresponds to the
coordinates on the stalk. Explicit form of metric tensor makes
possible the computation of the components of metric connection
for it but we do not need them in what follows. \par
     Much more interesting objects are the images of vector
fields on $TM$ under the action of maps $\pi$ and $\rho$ from
\thetag{3.5}. In general they could not be treated as vector
fields on $M$. They are the vector-valued functions whose
argument is the point of $TM$ and whose value is the vector
tangent to $M$ at the point being the projection of the argument.
Such function composes the algebra over the ring of scalar
functions on $TM$. This algebra can easily be expanded up to the
algebra of tensor-valued functions on $M$ with the argument from
$TM$. It is natural to call it the expanded algebra of tensor
fields on $M$. The first example of the element of such algebra
is the force field \thetag{2.6} for Newtonian dynamical system.
This field has the real physical meaning when the manifold $M$ is
the configuration space for some real mechanical system
restricted by inner bounds. \par
     Expanded algebra of tensor fields on $M$ is equipped with
the natural operations of tensor product and contraction.
Presence of metric on $M$ adds two operations of covariant
differentiation. First of them is the differentiation by velocity
or the velocity gradient. For scalar, vectorial and covector
fields it is defined by the following expressions
$$
\tilde\nabla_i\varphi=\frac{\partial\varphi}{\partial v^i}
\hskip 3em
\tilde\nabla_i X^m=\frac{\partial X^m}{\partial v^i}
\hskip 3em
\tilde\nabla_i X_m=\frac{\partial X_m}{\partial v^i} \tag3.9
$$
Second is the covariant differentiation by coordinate or the
space gradient. It is the modification of the ordinary covariant
differentiation. The result of its application to the scalar,
vectorial and covector fields is as follows
$$
\aligned
\nabla_i\varphi &=\frac{\partial\varphi}{\partial x^i}-
\varGamma^p_{ik}\frac{\partial\varphi}{\partial v^p}v^k\\
\nabla_i X^m &=\frac{\partial X^m}{\partial x^i}+
\varGamma^m_{ip} X^p-\varGamma^p_{ik}
\frac{\partial X^m}{\partial v^p}v^k\\
\nabla_i X_m &=\frac{\partial X_m}{\partial x^i}-
\varGamma^p_{im} X_p-\varGamma^p_{ik}
\frac{\partial X_m}{\partial v^p}v^k
\endaligned\tag3.10
$$
For other tensor fields the action of \thetag{3.9} and
\thetag{3.10} is continued according to the condition of
concordance with the operations of tensor product and
contraction. Ordinary tensor fields are the part of expanded
algebra. Velocity gradient for them is always zero while the
space gradient coincides with ordinary covariant derivative.
Particularly
$$
\nabla_k g_{ij}=0\hskip 10em \tilde\nabla_k g_{ij}=0 \tag3.11
$$
The relationships \thetag{3.11} express the compatibility of
metric and metrical connection in terms of the above covariant
derivatives. \par
     One more element of the expanded algebra is the vector field
of the velocity \thetag{2.4}. For the velocity and space
gradients of it we have
$$
\nabla_k v^i=0\hskip 10em \tilde\nabla_k v^i=\delta^i_k \tag3.12
$$
Scalar field of modulus of velocity is defined by $\bold v$
according to the following formula
$$
v^2=|\bold v|^2=g_{ij}v^iv^j \tag3.13
$$
For the gradients of the scalar field defined by \thetag{3.13}
we have
$$
\nabla_k v=0\hskip 10em \tilde\nabla_k v=N_k=g_{kq}N^q \tag3.14
$$
The quantities $N^q$ in \thetag{3.14} are the components of the
unit vector field $\bold N$ directed along the velocity $\bold v
= v \bold N$. Gradients for this vector field are the following
$$
\nabla_k \bold N^i=0\hskip 10em \tilde\nabla_k \bold N^i=v^{-1}
(\delta^i_k-N_kN^i) \tag3.15
$$
They are calculated on the base of \thetag{3.12} and
\thetag{3.14}. Components of the matrix $P^i_k=\delta^i_k-N_kN^i$
in \thetag{3.15} are the components of operator valued field
$\bold P$ of normal projectors on a hyperplane that is
perpendicular to $\bold v$. Covariant derivatives for $\bold P$
itself are
$$
\nabla_k P^i_j=0\hskip 10em \tilde\nabla_k P^i_j=-v^{-1}
(g_{jq} P^q_k N^i+N_j P^i_k)\tag3.16
$$
Along with $\bold P$ we define the additional projector-valued
field with the components $Q^i_j=N_jN^i$. For it we have $\bold P
+ \bold Q=\bold 1$. Covariant derivatives of $\bold Q$ are easily
calculated from \thetag{3.16}
$$
\nabla_k Q^i_j=0\hskip 10em \tilde\nabla_k Q^i_j=v^{-1}
(g_{jq} P^q_k N^i+N_j P^i_k) \tag3.17
$$
In addition to the properties \thetag{3.16} and \thetag{3.17} we
can see that projector-valued fields $\bold P$ and $\bold Q$ are
symmetric respective to the metric $g_{ij}$ on $M$. \par
     Let the functions $x^1(t),\dots,x^n(t)$ define the
parametric curve on $M$ in local map. The derivatives $\partial_t
x^i$ define the tangent vector to $M$ (it may be treated as
velocity vector of a point moving along this curve). Suppose that
at each point of this curve we have the tangent vector $\bold U$
to $M$ (but possibly not tangent to the curve) with the
components $U^i(t)$. In other words we have the vector-valued
function on the curve. Let us produce another vector-valued
function on that curve according to the formula
$$
\nabla_tU^i=\partial_tU^i+\varGamma^i_{jk}U^j\partial_tx^k
\tag3.18
$$
Formula \thetag{3.18} defines the covariant derivative of the
vector-valued function on the curve by the parameter $t$ of it.
It is well known that such derivative is zero by parallel
displacement of the vector along the curve.\par
     Let's add the functions $v^1(t),\dots,v^n(t)$ to $x^1(t),
\dots,x^n(t)$ and consider all them as the curve lifted from $M$
to $TM$. Such lifting is called natural if $v^i(t)=\partial_t
x^i(t)$. However here we consider arbitrary (may be not natural)
liftings. Let us define some vector field $\bold U$ of expanded
algebra in some neighborhood of the lifted curve. Substituting
$x^1(t),\dots,x^n(t)$, $v^1(t),\dots, v^n(t)$ for its argument
we obtain the vector-valued function on the former curve.
For this function we have
$$
\nabla_tU^i=\nabla_kU^i\partial_tx^k+\tilde\nabla_kU^i\nabla_tv^k
\tag3.19
$$
Formula \thetag{3.19} approves the names velocity and space
gradients for the covariant derivatives \thetag{3.9} and
\thetag{3.10}. It is easily modified for the case of scalar,
covectorial and all other types of tensor fields of expanded
algebra.\par
     Let us find the relation  between the ordinary covariant
derivatives on $TM$ and the modified covariant derivatives
\thetag{3.9} and \thetag{3.10} on the manifold $M$ itself. In
order to do it consider the pair of vector fields $\bold X$ and
$\bold Y$ on the tangent bundle TM. For the projections $\pi$ and
$\rho$ from \thetag{3.5} applied to the commutator of these vector
fields we derive
$$
\align
\split
\pi([\bold X, \bold Y])&=\nabla_{\pi(\bold X)}\pi(\bold Y)-
\nabla_{\pi(\bold Y)}\pi(\bold X)+\\
&\hphantom{=\nabla_{\pi(\bold X)}\pi(\bold Y)}
+\tilde\nabla_{\rho(\bold X)}
\pi(\bold Y)-\tilde\nabla_{\pi(\bold Y)}\pi(\bold X)
\endsplit
\tag3.20\\
\split
\rho([\bold X, \bold Y])&=\nabla_{\pi(\bold X)}\rho(\bold Y)-
\nabla_{\pi(\bold Y)}\rho(\bold X)+\\
&+\tilde\nabla_{\rho(\bold X)}
\rho(\bold Y)-\tilde\nabla_{\rho(\bold Y)}\rho(\bold X)-
R(\pi(\bold X),\pi(\bold Y)) \bold v
\endsplit
\tag3.21
\endalign
$$
Here $R(\bold A,\bold B)$ is the operator-valued skew-symmetric
bilinear form defined by the curvature tensor of $M$. Formulae
\thetag{3.20} and \thetag{3.21} are proved by the direct
calculations in the coordinates.\par
     Let $\bold Z$ be one more vector field on $TM$ and $\varphi$
be the scalar field on $TM$ which also may be treated as a scalar
field of the expanded algebra on $M$. Then
$$
\nabla_{\bold X}\varphi=\partial_{\bold X}\varphi=
\nabla_{\pi(\bold X)}\varphi+\tilde\nabla_{\rho(\bold X)}\varphi
\tag3.22
$$
Left hand side of \thetag{3.22} contain the ordinary covariant
derivative on $TM$ which coincides for the scalar function with
the derivative along the vector $X$. Covariant derivatives in the
right hand sides of \thetag{3.20}, \thetag{3.21} and
\thetag{3.22} are that of \thetag{3.9} and \thetag{3.10}. For to
calculate the covariant derivatives on $TM$ we use the following
formula from \cite{6}
$$
\aligned
2\tilde g(\nabla_{\bold X}\bold Y,\bold Z)&=\partial_{\bold X}
\tilde g(\bold Y,\bold Z)+\partial_{\bold Y}\tilde g(\bold X,
\bold Z)-\partial_{\bold Z}\tilde g(\bold X,\bold Y)+\\
&+\tilde g([\bold Z,\bold X],\bold Y)+\tilde g([\bold Z,
\bold Y],\bold X)+\tilde g([\bold X,\bold Y],\bold Z)
\endaligned\tag3.23
$$
Covariant derivative on $TM$ in \thetag{3.23} is ordinary one.
Now let us use the relationship
$$
\tilde g(\bold X,\bold Y)=g(\pi(\bold X),\pi(\bold Y))+
g(\rho(\bold X),\rho(\bold Y))\tag3.24
$$
The relationship \thetag{3.24} is derived directly from
\thetag{3.6}. As the result of substitution of \thetag{3.24} and
\thetag{3.20}, \thetag{3.21}, \thetag{3.22} into the identity
\thetag{3.23} we get two formulae
$$
\align
&\aligned
\pi(\nabla_{\bold X}\bold Y)&=\nabla_{\pi(\bold X)}\pi(\bold Y)+
\tilde\nabla_{\rho(\bold X)}\pi(\bold Y)-\\
&-\frac{1}{2}(R(\rho(\bold Y,
\bold v)\pi(\bold X)+R(\rho(\bold X),\bold v)\pi(\bold Y))
\endaligned\tag3.25\\
&\aligned
\rho(\nabla_{\bold X}\bold Y)=\nabla_{\pi(\bold X)}\rho(\bold Y)
&+\tilde\nabla_{\rho(\bold X)}\rho(\bold Y)-\\ &-\frac{1}{2}
R(\pi(\bold X),\pi(\bold Y))\bold v
\endaligned\tag3.26
\endalign
$$
Covariant derivatives in left hand sides of \thetag{3.25} and
\thetag{3.26} are ordinary ones like in \thetag{3.23}. Covariant
derivatives in right hand sides of \thetag{3.25} and
\thetag{3.26} are expanded ones. They should be treated as in
\thetag{3.9} and \thetag{3.10}.\par
\head
4. Variations of trajectories and the equations of weak
normality.
\endhead
     Let us consider the Newtonian dynamical system \thetag{2.2}.
In the second of the equations \thetag{2.2} we change the
ordinary derivative of the velocity by its covariant derivative
according to the formula \thetag{3.18}. This gives us the
equations
$$
\partial_tx^i=v^i \hskip 10em \nabla_tv^i=F^i\tag4.1
$$
containing the force vector of \thetag{2.5}. The equations
\thetag{4.1} are more natural form for the Newton's second law
on the manifold $M$.\par
     For the Newtonian dynamical system \thetag{4.1} now we
consider the Cauchy problem with the following initial data
$$
\left.x^i\right|_{t=0}=x^i(s)\hskip 10em
\left.\partial_tx^i\right|_{t=0}=v^i(s)\tag4.2
$$
depending on some parameter $s$. Because of \thetag{4.2} the
trajectories of the dynamical system also depend on $s$. In
local coordinates they are given by the functions $x^i=x^i(t,s)$.
Their derivatives by $s$ are the coordinates of some vector
$\boldsymbol\tau$ tangent to $M$. It is the vector of variations
of trajectories $\tau^i(t,s)=\partial_sx^i(t,s)$. The time
derivative of $\boldsymbol\tau$ is the vector of variations of
velocities
$$
\nabla_t\tau^i=\partial_t\tau^i+\varGamma^i_{jk}\tau^j v^k\tag4.3
$$
We shall use \thetag{4.3} to obtain the equations for the vector
$\boldsymbol\tau$ from the equations \thetag{4.1} of the
dynamical system itself. In order to do it let us differentiate
\thetag{4.1} by $s$ and combine the result with \thetag{4.3} and
the time derivative of \thetag{4.3}. Then we obtain
$$
\nabla_t\nabla_t\tau^i+R^i_{jkq}v^jv^k\tau^q=\nabla_qF^i\tau^q+
\tilde\nabla_qF^i\nabla_t\tau^q\tag4.4
$$
Here $R^i_{jqk}$ is the curvature tensor for $M$. So the
variation vector $\boldsymbol\tau$ satisfies the ordinary
differential equations of the second order \thetag{4.4}.\par
     The normal shift  condition according to \cite{1} and
\cite{2} consists in the orthogonality of $\boldsymbol\tau$ and
the vector of velocity $\bold v$. It is convenient to express
$\bold v$ via the unit vector $\bold N$ from \thetag{3.14} and
\thetag{3.15}. According to the method developed in \cite{1} and
\cite{2} we introduce the function $\varphi$ as a scalar product
$$
\varphi=\left<\boldsymbol\tau,\bold N \right>=
\tau^iN_i=g_{ij}\tau^iN^j \tag4.5
$$
For its time derivative by differentiating \thetag{4.5} we obtain
$$
\partial_t\varphi=g_{ij}\nabla_t\tau^iN^j+g_{ij}\tau^i\nabla_tN^j
$$
Because of \thetag{3.11} and \thetag{3.19} the metric tensor makes
no contribution by covariant differentiation. To calculate
$\nabla_tN^j$ we use the relationships \thetag{3.15} and
\thetag{3.19}. As a result we have
$$
\partial_t\varphi=g_{ij}\nabla_t\tau^iN^j+v^{-1}g_{ij}\tau^i
P^j_kF^k
$$
Because $\bold P$ is the symmetric operator field we may rewrite
this expression in the following form
$$
\partial_t\varphi=g_{ij}\nabla_t\tau^iN^j+v^{-1}g_{ij}F^iP^j_q
\tau^q\tag4.6
$$
Differentiating \thetag{4.6} by $t$ we obtain the second
derivative for $\varphi$
$$
\partial_{tt}\varphi=\partial_t(g_{ij}\nabla_t\tau^iN^j)+
\partial_t(v^{-1}g_{ij}F^iP^j_q\tau^q)\tag4.7
$$
Taking into account \thetag{3.15} and \thetag{3.19} we may write
the first summand in \thetag{4.7} as
$$
\partial_t(g_{ij}\nabla_t\tau^iN^j)=\nabla_{tt}\tau^iN_i+v^{-1}
F_iP^i_q\nabla_t\tau^q \tag4.8
$$
The first summand in \thetag{4.8} in turn is transformed by use
of the equation \thetag{4.4} for the vector of variation of
trajectory
$$
\nabla_{tt}\tau^iN_i=N_i\nabla_qF^i\tau^q+N_i\tilde\nabla_qF^i
\nabla_t\tau^q-N_iR^i_{\alpha q\beta}\tau^qv^\alpha v^\beta
\tag4.9
$$
Let us insert the projectors $\bold P$ and $\bold Q$ into all
terms in the right hand side of \thetag{4.9}. In order to do it
we use the decomposition of identical operator as $\bold 1 =
\bold P + \bold Q$. For the first summand we have
$$
N_i\nabla_qF^i\tau^q=\nabla_iF^kN_kP^i_q\tau^q+\nabla_qF^kN^qN_k
\tau^iN_i \tag4.10
$$
For the second summand in \thetag{4.9} by the same way we derive
$$
N_i\tilde\nabla_qF^i\nabla_t\tau^q=\tilde\nabla_iF^kN_kP^i_q
\nabla_t\tau^q+\tilde\nabla_qF^kN^qN_k\nabla_t\tau^iN_i \tag4.11
$$
The third summand in \thetag{4.9} vanishes because the vectors
$\bold N$ and $\bold v$ are collinear $\bold v=|\bold v|\bold N$.
Indeed
$$
N_iR^i_{\alpha q \beta}\tau^qv^\alpha v^\beta=|\bold v|^2
R_{i\alpha q\beta}N^iN^\alpha N^\beta\tau^q \tag4.12
$$
Curvature tensor is skew-symmetric in $i$ and $\alpha$. Therefore
the result of contraction in \thetag{4.12} is zero.\par
     Now  let us transform  the second summand in \thetag{4.7}
using the relationships \thetag{3.14} and \thetag{3.19}
$$
\partial(v^{-1}g_{ij}F^iP^j_k\tau^k)=-v^{-2}F^kN_kF_iP^i_q\tau^q+
v^{-1}F_iP^i_q\nabla_t\tau^q+v^{-1}\nabla_t(F_iP^i_q)\tau^q
\tag4.13
$$
Then let us transform the last summand in \thetag{4.13} with the
help of \thetag{3.16} and \thetag{3.19}
$$
\aligned
v^{-1}\nabla_t(F_iP^i_q)\tau^q&=\nabla_kF^iN^kP^i_q\tau^q+v^{-1}
\tilde\nabla_kF_iF^kP^i_q\tau^q-\\
&-v^{-2}F^kN_kF_iP^i_q\tau^q-
v^{-2}F_jP^j_kF^k\tau^iN_i
\endaligned\tag4.14
$$
As a result the second summand in \thetag{4.7} may be obtained by
substituting \thetag{4.14} into \thetag{4.13} and the first one
may be obtained by substitution of \thetag{4.10} and
\thetag{4.11} into \thetag{4.9} followed by substitution of
\thetag{4.9} into \thetag{4.8}. By analyzing the obtained
formulae \thetag{4.5}, \thetag{4.6} and \thetag{4.7} for
$\varphi$, $\partial_t\varphi$ and $\partial_{tt}\varphi$ we find
that all they are the linear functionals with respect to vectors
$boldsymbol\tau$ and $\nabla_t\boldsymbol\tau$ forming the phase
space for the integral trajectories of the system of differential
equations \thetag{4.4}. Using $\varphi$, $\partial_t\varphi$ and
$\partial_{tt}\varphi$ we construct another functional $L$ of
the form
$$
\partial_{tt}\varphi-P\partial_t\varphi-Q\varphi=L \tag4.15
$$
The coefficients for $L$ in the formula \thetag{4.15} we
define as
$$
P=\tilde\nabla_qF^kN^qN_k\hskip 10em Q=\tilde\nabla_qF^kN^qN_k-
v^{-2}P^q_kF_qF^k
$$
Then the functional $L$ itself may be written in the form
$$
L=\xi_iP^i_q\nabla_t\tau^q+\zeta_iP^i_q\tau^q \tag4.16
$$
The coefficients $\xi_i$ and $\zeta_i$ in \thetag{4.16} are
defined by the above calculations. They are the following
$$
\align
&\xi_i=\tilde\nabla_iF^kN_k+2 v^{-1}F_i\tag4.17\\
&\aligned
\zeta_i=(\nabla_iF_k&+\nabla_kF_i-2 v^{-2}F_iF_k)N^k+\\
&+v^{-1}(\tilde\nabla_kF_iF^k-\tilde\nabla_kF^qN^kN_qF_i)
\endaligned\tag4.18
\endalign
$$
For the dynamical system \thetag{4.1} to accept the normal shift
on $M$ (see \cite{1} and \cite{2}) the functional $L$ should
identically vanish. This condition in \cite{1} and \cite{2} is
called the condition of weak normality. In the present situation
it gives us the following equations for $\xi_i$ and $\zeta_i$
from \thetag{4.17} and \thetag{4.18}
$$
\xi_iP^i_q=0\hskip 10em \zeta_iP^i_q=0 \tag4.20
$$
The equations \thetag{4.20} may be rewritten in the following
explicit form
$$
\align
&(v^{-1}F_i+\tilde\nabla_i(F^kN_k))P^i_q=0\tag4.21\\
&\aligned
(\nabla_iF_k+\nabla_kF_i&-2 v^{-2}F_iF_k)N^kP^i_q+\\
v^{-1} (\tilde\nabla_kF_iF^k&-\tilde\nabla_kF^rN^kN_rF_i)
P^i_q=0\endaligned
\tag4.22
\endalign
$$
The total number of the equations \thetag{4.21} and \thetag{4.22}
is $2n$. It coincides with the twiced dimension of the manifold
$M$. However because of presence of projector matrices $P^i_q$
in them they are not independent. The number of independent
equations in the system of \thetag{4.21} and \thetag{4.22} is
$2n-1$ which is in concordance with the results of \cite{1} and
\cite{2}.\par
     The equations \thetag{4.21} and \thetag{4.22} are the
covariant form of the equations of weak normality on an arbitrary
Riemannian manifold. In Euclidean case $M=\Bbb R^n$ they were
derived in \cite{1} and \cite{2} by use of spherical coordinates
in the space of velocities. The question of introducing the proper
spherical coordinates here in the general situation is
interesting but it is the subject for separate paper. Analyzing
the equations \thetag{4.21} one can see that if the vector of
force is decomposed into two parts first being along the velocity
and second being perpendicular to it then the second part is
defined by the first one. This fact was observed in \cite{1} and
\cite{2}. It remains true for the general non-Euclidean
situation.\par
     This paper is written under the financial support of two of
authors (Boldin A.Yu. and Sharipov R.A.) by Russian Fund for
Fundamental Researches.\par

\enddocument

\Refs
\ref\no 1\by Boldin A.Yu., Sharipov R.A. \paper Dynamical systems
accepting the normal shift. \jour Theor. and Math. Phys. \yr 1993
\vol 97\issue 3\pages 386--395\lang Russian
\endref
\ref\no 2\by Boldin A.Yu., Sharipov R.A. \paper Dynamical systems
accepting the normal shift. \jour Dikladi Akade\-mii Nauk.
\yr 1994 \vol 334 \issue 2 \pages 165--167\lang Russian
\endref
\ref\no 3\by Bianchi L.\paper Sopra le deformazioni isogonali
delle superficie a curvatura constante in geometria elliptica ed
hiperbolica. \yr 1911 \jour  Annali di Matem.\vol 18\issue 3
\pages 185--243
\endref
\ref\no 4\by Tenenblat K. \paper B\"acklund theorem for
submanifolds of space forms and a generalized wave equation.
\yr 1985 \jour  Bol. Soc. Bras. Math.\vol 18\issue 2\pages 67--92
\endref
\ref\no 5\by Sharipov R.A.\paper Problem of metrizability for the
dynamical systems accepting the normal shift. \jour Theor. and Math.
Phys. \toappear\lang Russain
\endref
\ref\no 6\by Kobayashy Sh., Nomidzu K.\book Foundations of
differential geometry. \vol 1\publ Intersc. Publ., New York -
London\yr 1963
\endref
\endRefs
\bye